\documentclass[a4paper,12pt]{article}
\usepackage{amssymb}
\usepackage{amsmath}
\usepackage{epsfig}
\usepackage{stmaryrd}
\usepackage{graphics}

\usepackage{latexsym}
\usepackage{rotating}
\title
{
 A counterpart of the
 WKI
 soliton
 hierarchy associated with
 $\textrm{so}(3,\mathbb{R})$
 }
\author{{
\normalsize
 $^1$Wen-Xiu Ma\thanks{Email: mawx@cas.usf.edu}$\ $, $^1$Solomon  Manukure  and $^2$Hong-Chan Zheng}\\
{\small $^1$Department of Mathematics and Statistics, University of South Florida, }\\
{\small Tampa, FL 33620-5700,
USA}
\\ {\small $^2$Department of Applied Mathematics, Northwestern Polytechnical University, }
\\
{\small 710072 Xi'an, PR China
}
}

\date{\nonumber}
\begin{document}
\maketitle
\date{\nonumber}

\newcommand{\R}{\mathbb{R}}

\numberwithin{equation}{section}

\vskip 2mm

\begin{abstract}

A counterpart of the Wadati-Konno-Ichikawa (WKI) soliton hierarchy,
associated with $\textrm{so}(3,\mathbb{R})$,
is
presented through the zero curvature formulation. Its spectral matrix is defined by the same linear combination of basis vectors as the WKI one, and
its Hamiltonian structures yielding Liouville integrability are furnished by the trace identity.

\vskip 1mm


\end{abstract}

\newtheorem{thm}{Theorem}[section]
\newtheorem{Le}{Lemma}[section]

\setlength{\baselineskip}{19.5pt}
\def \part {\partial}
\def \be {\begin{equation}}
\def \ee {\end{equation}}
\def \bea {\begin{eqnarray}}
\def \eea {\end{eqnarray}}
\def \ba {\begin{array}}
\def \ea {\end{array}}
\def \si {\sigma}
\def \al {\alpha}
\def \la {\lambda}
\def \D {\displaystyle}

\section{Introduction}

Soliton hierarchies consist of commuting nonlinear partial differential equations with Hamiltonian structures, and they are usually generated from given spectral problems associated with matrix Lie algebras (see, e.g., \cite{AblowitzC-book1991}-\cite{Drinfel'dS-SMD1981}). Typical examples include the Korteweg-de Vries hierarchy \cite{Lax-CPAM1968}, the Ablowitz-Kaup-Newell-Segur hierarchy \cite{AblowitzKNS-SAM1974}, the Kaup-Newell hierarchy \cite{KaupN-JMP1978} and the Wadati-Konno-Ichikawa (WKI) hierarchy \cite{WadatiKI-JPSJ1979}.

When associated matrix Lie algebras are semisimple, the trace identity can be used to construct Hamiltonian structures of soliton hierarchies
\cite{Tu-JPA1989,Ma-CAMA1992}.
When
associated matrix Lie algebras
are
non-semisimple,
we obtain
 integrable couplings \cite{MaXZ-PLA2006,MaXZ-JMP2006}, and
the variational identity provides a basic technique to generate
 their Hamiltonian structures \cite{MaC-JPA2006,Ma-JPA2007}.
Usually, the existence of
  bi-Hamiltonian structures \cite{Magri-JMP1978} implies Liouville integrability,
  often generating hereditary recursion operators (see, e.g., \cite{FuchssteinerF-PD19812,Olver-book1986,CasatiDO-JGP2008}).
The most widely used 3-dimensional simple Lie algebra in soliton theory is
the special linear Lie algebra
sl$(2,\mathbb{R})$. We would like to use the other 3-dimensional simple Lie algebra, the special orthogonal Lie algebra so$(3,\mathbb{R})$. Those two Lie algebras are only the two real 3-dimensional Lie algebras, whose derived algebras are 3-dimensional, too.

Let us briefly outline the steps of our procedure to construct soliton hierarchies by the zero curvature formulation (see, e.g., \cite{Tu-JPA1989,Ma-CAMA1992} for details).

{\bf Step 1 - Introducing a spatial spectral problem:}

Take a matrix loop algebra $  \tilde {\mathfrak {g}} $, associated with a given matrix Lie algebra ${\mathfrak {g}} $,
 often being semisimple.
Then, introduce
a spatial spectral
problem
\be \phi_x=U\phi,\ U=U(u,\lambda)\in  \tilde {\mathfrak {g}},
\label{eq:ssp:ma-3rd-sh-so(3,R)}
\ee
where $u$ denotes a column dependent variable and $\lambda $ is the spectral parameter.

{\bf Step 2 - Computing zero curvature equations:}

We search for a solution of the form
\be W=W(u,\lambda)=\sum_{i\ge 0} W_{0,i}\lambda ^{-i},\ W_{0,i}\in \mathfrak {g} , \ i\ge 0, \label{eq:Wofgis:ma-3rd-sh-so(3,R)} \ee
to the stationary zero curvature equation
\be W_x=[U,W]. \label{eq:SZCEofgis:ma-3rd-sh-so(3,R)}\ee
Then, use this solution $W$ to introduce
the Lax matrices
\be V^{[m]} =V^{[m]}(u,\lambda )=  (\lambda ^m W)_+ +\Delta _m\in \tilde {\mathfrak {g}} ,\ m\ge 0,
\label{eq:V^{[m]}ofgis:ma-3rd-sh-so(3,R)}
\ee
where $P_+$ denotes the polynomial part of $P$ in $\lambda $,
and formulate
the temporal
spectral problems
\be \phi_{t_m} = V^{[m]}\phi=V^{[m]}(u,\lambda )\phi, \ m\ge 0.\label{eq:tsp:ma-3rd-sh-so(3,R)} \ee
The crucial point is to input the modification terms: $\Delta_{m}\in \tilde {\mathfrak {g}},\ m\ge 0$,
which aims to guarantee that the compatibility conditions of \eqref{eq:ssp:ma-3rd-sh-so(3,R)} and \eqref{eq:tsp:ma-3rd-sh-so(3,R)},
i.e.,
the zero curvature equations
\be U_{t_m}-V^{[m]}_x+[U,V^{[m]}]=0,\ m\ge 0,\ee
will generate soliton equations.
We write
the resulting hierarchy of soliton equations of evolution type as follows:
\be u_{t_m}=K_m(u),\ m\ge 0. \label{eq:givenSH:ma-3rd-sh-so(3,R)}  \ee

{\bf Step 3 - Constructing Hamiltonian structures:}

Compute
 Hamiltonian functionals ${\cal H}_m$'s by applying the trace identity \cite{Tu-JPA1989,Ma-CAMA1992}:
\be  \frac \delta {\delta u}\int \textrm{tr}( \frac {\partial U}{\partial \lambda } W)\, dx= \lambda ^{-\gamma } \frac \partial {\partial \lambda }
\lambda ^\gamma \,\textrm{tr}( \frac {\partial U}{\partial u } W ),\ \gamma=-\frac \lambda 2 \frac d {d\lambda } \ln |\textrm{tr}(W^2)|,
\label{eq:traceid:ma-3rd-sh-so(3,R)}
\ee
or more generally, the variational identity \cite{MaC-JPA2006,Ma-book2010}:
\be \frac \delta {\delta u}\int \langle \frac {\partial U}{\partial \lambda }, W\rangle\, dx= \lambda ^{-\gamma } \frac \partial {\partial \lambda }
\lambda ^\gamma \,\langle \frac {\partial U}{\partial u }, W \rangle,\ \gamma=-\frac \lambda 2 \frac d {d\lambda } \ln |\langle W, W \rangle |,
\label{eq:vid:ma-3rd-sh-so(3,R)}
\ee
where
$\langle \cdot ,\cdot \rangle$ is a non-degenerate, symmetric and ad-invariant bilinear form on the underlying matrix loop algebra $ \tilde {\mathfrak {g}} $. Then, construct Hamiltonian structures for the whole hierarchy \eqref{eq:givenSH:ma-3rd-sh-so(3,R)}:
\be u_{t_m}=K_m(u)=J\frac {\delta {\cal H}_m}{\delta u},\ m\ge 0. \ee
The generating functional
$\int \textrm{tr}(\frac {\partial U}{\partial \lambda } W)\,dx$ or $\int \langle \frac {\partial U}{\partial \lambda }, W\rangle\,dx
$
will be used to generate the Hamiltonian functionals $\{{\cal H}_m\}_0^\infty $ in the above Hamiltonian structures.
Usually, the recursion structure of a soliton hierarchy leads to its bi-Hamiltonian structures and Liouville integrability.

In this paper, starting from the 3-dimensional special orthogonal Lie algebra so$(3,\mathbb{R})$,
we would like to present a counterpart of the WKI soliton hierarchy. The counterpart soliton hierarchy consists of commuting bi-Hamiltonian evolution equations,
which are of differential function type but not of differential polynomial type,
and its
 corresponding Hamiltonian structures will be furnished by the trace identity. Therefore,
 all equations in the counterpart soliton hierarchy provide a new example of soliton hierarchies
 associated with so$(3,\mathbb{R})$ (see \cite{Ma-AMC2013,Ma-JMP2013} for two examples of Ablowitz-Kaup-Newell-Segur and Kaup-Newell types).
A few concluding remarks will be given in the final section.

\section{A counterpart of the WKI soliton hierarchy}

\subsection{The WKI hierarchy}

Let us recall the WKI soliton hierarchy \cite{WadatiKI-JPSJ1979,BoitiPT-PTP1983} for comparison's sake. Its
corresponding
special matrix reads
\be
U=U(u,\lambda )= \lambda e_1+\lambda p e_2+\lambda q e_3,
\ee
where $e_1,e_2$ and $e_3$, forming
 a basis of the special linear Lie algebra sl$(2,\mathbb{R})$, are defined as follows:
\be e_1=\left[
\ba {ccc}  1&0 \vspace{2mm}\\
0& -1   \ea
 \right],\
e_2=\left[
\ba {ccc}  0&1 \vspace{2mm}\\
0& 0  \ea
 \right],\
e_3=\left[
\ba {ccc}  0&0 \vspace{2mm}\\
1& 0  \ea
 \right],
\ee
whose commutator relations are
\[ [e_1,e_2]=2 e_2,\ [e_1,e_3]=-2 e_3,\ [e_2,e_3]=e_1. \]

A solution of the form
\be W
=aU+b_xe_2+c_xe_3
=\lambda a e_1+(\lambda p a +b_x)e_2+
(\lambda q a +c_x) e_3 \ee
to the stationary zero curvature equation \eqref{eq:SZCEofgis:ma-3rd-sh-so(3,R)} is determined by
\be  \left\{ \ba {l}
a_x=pc_x-qb_x,
\vspace{2mm}\\
\lambda (pa)_x+b_{xx}=2\lambda b_x,
\vspace{2mm}\\
\lambda (qa)_x+c_{xx}=-2\lambda c_x.
\ea \right. \label{eq:RRofSZCEforWKI:ma-3rd-sh-so(3,R)} \ee
Upon setting
 \begin{equation}
a =\sum _{i\ge 0}a_i\lambda ^{-i},
\ b=  \sum _{i\ge 0}b_i\lambda ^{-i},
 \ c=  \sum _{i\ge 0}c_i\lambda ^{-i},
\  i\ge 0 ,
\label{eq:ConcretedFormulaofWforWKI:ma-3rd-sh-so(3,R)}\end{equation}
and
 choosing
the initial values
\be a_0=\frac 1 {\sqrt{pq+1}},\  b_0=\frac p  {2\sqrt{pq+1}},\ c_0=-\frac q  {2\sqrt{pq+1}},\ee
the system \eqref{eq:RRofSZCEforWKI:ma-3rd-sh-so(3,R)} recursively defines the infinite sequence of $\{a_i,b_i,c_i|\, i\ge 1\}$ as follows:
\begin{equation}
\left[ \ba {c}
-c_{i+1}\vspace{2mm}\\
b_{i+1}\ea \right]
=\Psi
\left[ \ba {c}
-c_{i}\vspace{2mm}\\
b_{i}\ea \right],\
\Psi=\left[ \ba {cc}
-\frac 12 \partial +\frac 14 \bar q \partial ^{-1} \bar p \partial ^2 & -\frac 14 \bar q \partial ^{-1}\bar q \partial ^2
\vspace{2mm}\\
\frac 14 \bar p \partial ^{-1}\bar p\partial ^2 & \frac 12\partial  -\frac 14 \bar p \partial ^{-1} \bar q \partial ^2
\ea \right]
,\ i\ge 0,
\label{eq:RR1_iofSZCEforWKI:ma-3rd-sh-so(3,R)}\end{equation}
and \begin{equation}
\vspace{2mm}\\
a_{i+1,x}= p c_{i+1,x}- q b_{i+1,x},
\ i\ge 0,\label{eq:RR2_iofSZCEforWKI:ma-3rd-sh-so(3,R)}
\end{equation}
with $\bar p$ and $\bar q$ being given by
\be \bar p= \frac p {\sqrt{pq+1}},\ \bar q= \frac q {\sqrt{pq+1}}. \ee
We impose the conditions on constants of integration:
\[ a_i|_{u=0}=b_i|_{u=0}=c_i|_{u=0}=0,\ i\ge 1 ,\]
which guarantee the uniqueness of the infinite sequence of $\{a_i,b_i,c_i|\, i\ge 1\}$.
So, the first two sets can be computed as follows:
\begin{equation}
  \ba {l}
  a_1 = \frac {pq_x-qp_x}{4(pq+1)^{\frac 32}},
  \
  b_1 =\frac {p_x}{4(pq+1)^{\frac 32}}  , \
 c_1 = \frac {q_x}{4(pq+1)^{\frac 32}};
\vspace{2mm}\\
 a_2 = \frac 1 {32(pq+1)^{\frac 72}} [
 5q^2p_x^2+(14pq+4)p_xq_x+5p^2q_x^2 \vspace{2mm}\\
 \qquad -4q(pq+1)p_{xx}-4p(pq+1)q_{xx}
 ],
 \vspace{2mm}\\
b_2=
-\frac 1 {64(pq+1)^{\frac 72}} [
 q(7pq+12)p_x^2-2p(pq-4)p_xq_x-5p^3q_x^2 \vspace{2mm}\\
 \qquad -4(pq+1)(pq+2)p_{xx}+4p^2(pq+1)q_{xx}
 ],
\vspace{2mm}\\
 c_2=
 -\frac 1 {64(pq+1)^{\frac 72}} [
 5q^3p_x^2+2q(pq-4)p_xq_x-p(7pq+12)q_x^2 \vspace{2mm}\\
 \qquad -4q^2(pq+1)p_{xx}+4(pq+1)(pq+2)q_{xx}
 ]
  .
\ea
 \nonumber
\end{equation}

Finally, upon taking
\be V^{[m]}= \lambda [(\lambda^m a)_+ U+(\lambda ^m b_x)_{+}e_2+(\lambda ^m c_x)_{+}e_3], \ m \ge 0,\ee
the corresponding zero curvature equations
\begin{equation}
U_{t_m} -V^{[m]}_x +[U, V^{[m]}] = 0, \ m\ge 0,
\label{eq:ZCEforshforWKI:ma-3rd-sh-so(3,R)}
\end{equation}
present the WKI hierarchy of commuting Hamiltonian equations:
\begin{equation}
u_{t_m}= K_m=
\left[\ba {c}
b_{m,xx}
\vspace{2mm} \\
c_{m,xx}\ea
 \right]=
 J
 \left[\ba {c}
-c_{m}
\vspace{2mm} \\
b_{m}\ea
 \right]
 =
 J \frac{\delta {\cal H}_{m}}{\delta  u}
 ,
\ m \ge 0,\label{eq:WKIsh:ma-3rd-sh-so(3,R)}
\end{equation}
with the Hamiltonian operator $J$ being defined by
\be J=\left[ \begin{matrix}
0  & \partial ^2 \vspace{2mm}
\\ -\partial ^2 & 0
\end{matrix}\right],\label{eq:JforWKI:ma-3rd-sh-so(3,R)} \ee
and the Hamiltonian functionals ${\cal H}_{m}$'s, by
\be
{\cal H}_{0}=\int 2\sqrt{pq+1} \, dx,\
{\cal H}_{1}=\int \frac {qp_x-pq_x}{4\sqrt{pq+1}(\sqrt{pq+1}+1)} \, dx, \label{eq:1H_mforWKI:ma-3rd-sh-so(3,R)} \ee
and
\be
{\cal H}_{m+1}=\int\bigl[-
\frac {2(pq+1)a_{m+1}+pc_{m,x}+qb_{m,x}}{2m}\bigr]\, dx ,\ m\ge 1.\label{eq:2H_mforWKI:ma-3rd-sh-so(3,R)}
\ee
The above Hamiltonian functionals ${\cal H}_{m}$'s, $m\ne 1$, can be worked out by
the trace identity \eqref{eq:traceid:ma-3rd-sh-so(3,R)}
with
\[
\ba {l}\textrm{tr}(W\frac {\part U}{\part \lambda} )= 2\lambda (pq+1)a+pc_x+qb_x,\vspace{2mm}\\
\textrm{tr}(W\frac {\part U}{\part p} )=\lambda (\lambda qa+c_x)=-2\lambda ^2 c,
\vspace{2mm}
\\
 \textrm{tr}(W\frac {\part U}{\part q} )=
 \lambda (\lambda p a+b_x)=
 2\lambda ^2 b,
 \ea
\]
and ${\cal H}_1$ can be computed directly from $(-c_1,b_1)^T$.

We point out that a generalized WKI soliton hierarchy was presented in \cite{Xu-PLA2002} and its binary nonlinearization was carried out in \cite{Xu-CSF2003}.
 A multi-component WKI hierarchy, and a multi-component generalized WKI hierarchy and their integrable couplings, were also analyzed in \cite{YaoZ-CSF2005} and \cite{XiaYC-CSF2005}, respectively.

\subsection{A counterpart of the WKI hierarchy}

We will make use of the 3-dimensional special orthogonal Lie algebra
$\textrm{so}(3,\mathbb{R})$,
 consisting of  $3\times 3 $ skew-symmetric real matrices.
 This Lie algebra is simple and
has the basis
\be e_1=\left[
\ba {ccc}  0&0&-1 \vspace{2mm}\\
0& 0& 0 \vspace{2mm}\\
1 & 0 &0  \ea
 \right],\
e_2=\left[
\ba {ccc}  0&0&0 \vspace{2mm}\\
0& 0& -1 \vspace{2mm}\\
0 & 1 &0  \ea
 \right],\
e_3=\left[
\ba {ccc}  0&-1&0 \vspace{2mm}\\
1& 0& 0 \vspace{2mm}\\
0 & 0 &0  \ea
 \right],
\ee
whose commutator relations are
\[ [e_1,e_2]=e_3,\ [e_2,e_3]=e_1,\ [e_3,e_1]=e_2. \]
Its derived algebra is itself, and so, 3-dimensional, too.
The corresponding matrix loop algebra we will use is
\be \widetilde {\textrm{so}}(3,\mathbb{R}) =
 \Bigl\{\,
\sum_{i\ge 0}M_i\lambda ^{n-i}
\,\bigl|\bigr.\, M_i\in \textrm{so}(3,\mathbb{R}),\ i\ge 0,\ n\in \mathbb{Z}   \Bigr\}.  \ee
The loop algebra $\widetilde {\textrm{so}}(3,\mathbb{R})$ contains matrices of the form
\[\lambda ^m e_1+\lambda ^n e_2+\lambda ^l e_3\]
 with arbitrary integers $m,n,l$, and it provides
a good starting point to generate soliton equations.

Let us now introduce
a spectral matrix
\be
U=U(u,\lambda)=\lambda e_1+\lambda p e_2+\lambda q e_3=
\left[\begin{matrix}
0& -\lambda q& -\lambda
 \vspace{1mm} \\
\lambda q &0 & -\lambda p\vspace{1mm}\\
\lambda  & \lambda p &0
\end{matrix}\right] \in \widetilde {\textrm{so}}(3,\mathbb{R}),\
u = \left[\begin{matrix} p \vspace{1mm} \\
q
\end{matrix}\right],
\label{eq:defofU:ma-3rd-sh-so(3,R)}
\ee
to formulate
a matrix spatial spectral problem
\begin{equation}
 \phi_x =U\phi =U(u,\lambda)\phi,\
  \phi = ( \phi_1,
\phi_2
,
\phi_3)^T
.\label{eq:SP:ma-3rd-sh-so(3,R)}
\end{equation}
The spectral matrix above is defined by the same linear combination of basis vectors as the WKI one \cite{WadatiKI-JPSJ1979}, but its underlying loop algebra is $\widetilde{\textrm{so}}(3,\mathbb{R})$, not isomorphic to $\widetilde{\textrm{sl}}(2,\mathbb{R})$.
The other two examples associated with $\widetilde{\textrm{so}}(3,\mathbb{R})$, as counterpart hierarchies of the Ablowitz-Kaup-Newell-Segur hierarchy and the Kaup-Newell hierarchy, were previously presented in
 \cite{Ma-AMC2013} and \cite{Ma-JMP2013}, respectively.

Then, we solve the stationary zero
curvature equation \eqref{eq:SZCEofgis:ma-3rd-sh-so(3,R)}, and it
becomes
\begin{equation}
\left\{ \ba {l} a_x=  pc_x-  qb_x,\vspace{2mm}\\
 \lambda (pa)_x + b_{xx}=-\lambda c_x
,\vspace{2mm}\\
\lambda (qa)_x+ c_{xx}=\lambda  b_x ,\ea \right. \label{eq:RRofSZCE:ma-3rd-sh-so(3,R)}
\end{equation}
if $W$ is chosen as
\begin{eqnarray}
&&
W =
aU+b_xe_2+c_xe_3
\nonumber\\
&&
\quad\  =
\lambda a e_1+(\lambda pa +b_x) e_2 + (\lambda qa+c_x) e_3
\nonumber
\vspace{2mm}
\\
&&
\quad \ =
\left[\begin{matrix}
0&- (\lambda qa+c_x)&   -\lambda a \vspace{1mm}
\\ \lambda qa+c_x &0&-(\lambda pa +b_x)
\vspace{1mm}\\
\lambda a &\lambda pa +b_x  &0
\end{matrix}\right]\in \widetilde {\textrm{so}}(3,\mathbb{R}).
\label{eq:W:ma-3rd-sh-so(3,R)}\end{eqnarray}

Further, we set
 \begin{equation}
a =\sum _{i\ge 0}a_i\lambda ^{-i},
\ b=  \sum _{i\ge 0}b_i\lambda ^{-i},
 \ c=  \sum _{i\ge 0}c_i\lambda ^{-i},
\  i\ge 0 ,
\label{eq:ConcretedFormulaofW:ma-3rd-sh-so(3,R)}\end{equation}
and
 take
the initial values
\be a_0=\frac 1 {\sqrt{p^2+q^2+1}},\  b_0=\frac q  {\sqrt{p^2+q^2+1}},\ c_0=-\frac p  {\sqrt{p^2+q^2+1}},\ee
which are required by
\[ a_{0,x}=pc_{0,x}-qb_{0,x},\ pa_0=-c_0,\ qa_0=b_0. \]
The system \eqref{eq:RRofSZCE:ma-3rd-sh-so(3,R)} then leads to
the following two recursion relations:
\begin{equation}
\left[ \ba {c}
c_{i+1}\vspace{2mm}\\
-b_{i+1}\ea \right]
=\Psi
\left[ \ba {c}
c_{i}\vspace{2mm}\\
-b_{i}\ea \right],\
\Psi=\left[ \ba {cc}
\tilde p \partial ^{-1} \tilde q \partial ^2 & \partial -\tilde p \partial ^{-1}\tilde p \partial ^2
\vspace{2mm}\\
-\partial +\tilde q \partial ^{-1}\tilde q \partial ^2 & -\tilde q \partial ^{-1} \tilde p \partial ^2
\ea \right]
,\ i\ge 0,
\label{eq:RR1_iofSZCE:ma-3rd-sh-so(3,R)}\end{equation}
and \begin{equation}
\vspace{2mm}\\
a_{i+1,x}= p c_{i+1,x}- q b_{i+1,x},
\ i\ge 0,\label{eq:RR2_iofSZCE:ma-3rd-sh-so(3,R)}
\end{equation}
with $\tilde p$ and $\tilde q$ being defined by
\be \tilde p= \frac p {\sqrt{p^2+q^2+1}},\ \tilde q= \frac q {\sqrt{p^2+q^2+1}}. \ee
We will show that all vectors $(c_i,-b_i)^T$, $i\ge 0$, are gradient and the adjoint operator of $\Psi$ is hereditary in the next section.
To determine the sequence of $\{a_i,b_i,c_i|\, i\ge 1\}$ uniquely,
we impose the following conditions on constants of integration:
 \begin{equation}
 a_i|_{u=0}=b_i|_{u=0}=c_i|_{u=0}=0,\ i\ge 1 .
 \label{eq:ZeroConstofIntofSZCE:ma-3rd-sh-so(3,R)}
 \end{equation}
This way, the first two sets can be computed as follows:
\begin{equation}
  \ba {l}
  a_1 = \frac {qp_x-pq_x}{(p^2+q^2+1)^{\frac 32}},
  \
  b_1 =-\frac {p_x}{(p^2+q^2+1)^{\frac 32}}  , \
 c_1 = -\frac {q_x}{(p^2+q^2+1)^{\frac 32}};
\vspace{2mm}\\
 a_2 =
 -\frac 1 {(p^2+q^2+1)^{\frac 72}}[
 (3p^2+\frac 12 q^2 +\frac 12 )p_x^2
 +5pqp_xq_x
 +(\frac 12 p^2+3q^2 +\frac 12 )q_x^2
 \vspace{2mm}\\
 \qquad
 -p(p^2+q^2+1)p_{xx}
 -q(p^2+q^2+1)q_{xx}
 ]
 ,
 \vspace{2mm}\\
b_2=
\frac 1 {(p^2+q^2+1)^{\frac 72}}[
 -\frac 12 q(6p^2+ q^2 + 1 )p_x^2
 +p(3p^2-2q^2+3)p_xq_x
 +\frac 52q(p^2 +1)q_x^2
 \vspace{2mm}\\
\qquad  +pq(p^2+q^2+1)p_{xx}
 -(p^2+1)(p^2+q^2+1)q_{xx}
 ]
,
\vspace{2mm}\\
 c_2=
 \frac 1 {(p^2+q^2+1)^{\frac 72}}[
 -\frac 52 p(q^2 +1 )p_x^2
 -q(3q^2-2p^2+3)p_xq_x
 +\frac 12 p(p^2 +6q^2+1)q_x^2
 \vspace{2mm}\\
\qquad
 +(q^2+1)(p^2+q^2+1)p_{xx}
 -pq(p^2+q^2+1)q_{xx}
 ]
  .
\ea
 \nonumber
\end{equation}

Let us explain how to derive the
recursion relations in \eqref{eq:RR1_iofSZCE:ma-3rd-sh-so(3,R)}.
First from \eqref{eq:RRofSZCE:ma-3rd-sh-so(3,R)}, we have
\[ \ba {l} a_{i,x}=pc_{i,x}-qb_{i,x}
 \vspace{2mm}\\
\ \  \quad = p[-(pa_i)_x-b_{i-1,xx}]-q[(qa_i)_x+c_{i-1,xx}]
 \vspace{2mm}\\
 \ \  \quad =-(p^2+q^2)a_{i,x}-\frac 12 (p^2+q^2)_xa_i-pb_{i-1,xx}-qc_{i-1,xx},\ i\ge 1.
 \ea\]
 This is equivalent to
\[ \sqrt{p^2+q^2+1} \,(\sqrt{p^2+q^2+1}\,a_i)_x=
-pb_{i-1,xx}-qc_{i-1,xx},\ i\ge 1,\]
which leads to
\begin{equation} a_i=-\frac 1 {\sqrt{p^2+q^2+1}} (
\partial ^{-1}\tilde p \partial ^2 b_{i-1}+
\partial ^{-1}\tilde q \partial ^2 c_{i-1}
),\ i\ge 1. \label{eq:RRfora_iofSZCE:ma-3rd-sh-so(3,R)}
 \end{equation}
Then
again by \eqref{eq:RRofSZCE:ma-3rd-sh-so(3,R)} and using
\eqref{eq:ZeroConstofIntofSZCE:ma-3rd-sh-so(3,R)},
we see that
 \begin{equation}
 c_{i+1}=  -pa_{i+1}-b_{i,x},\ b_{i+1}= qa_{i+1}+c_{i,x},\ i\ge 0.
 \label{eq:RR3ofSZCE:ma-3rd-sh-so(3,R)}
 \end{equation}
Now the recursion relations in \eqref{eq:RR1_iofSZCE:ma-3rd-sh-so(3,R)} follows from the above recursion relation \eqref{eq:RRfora_iofSZCE:ma-3rd-sh-so(3,R)} for $a_i$.

Note that the first three sets of $\{a_i,b_i,c_i|\, i\ge 1\}$ are of differential function type.
This is actually true for all sets. We prove here that the whole sequence of $\{a_i,b_i,c_i|\, i\ge 1\}$ is of differential function type.
First from the stationary zero
curvature equation \eqref{eq:SZCEofgis:ma-3rd-sh-so(3,R)}, we can compute
\[ \frac d {dx}\textrm{tr}(W^2)=2\textrm{tr}(WW_x)
=2\textrm{tr}(W[U,W])=0.
 \]
 Thus, by \eqref{eq:ZeroConstofIntofSZCE:ma-3rd-sh-so(3,R)}, we obtain an equality
 \[ (p^2+q^2+1)a^2\lambda ^2 +2a(pb_x+qc_x)\lambda +b_x^2+c_x^2 =\lambda ^2 ,\]
 since for $W$ defined by \eqref{eq:W:ma-3rd-sh-so(3,R)}, we have
  \[ \frac 12 \textrm{tr}(W^2)=-(p^2+q^2+1)a^2\lambda ^2 -2a(pb_x+qc_x)\lambda -b_x^2-c_x^2 .\]
This equality gives
a formula to define $a_{i+1}$ by using the previous sets $\{a_j,b_j,c_j|\, j\le i\}$:
\[ \ba{l}
\D a_{i+1}=
-\frac 1 {2a_0}
\bigl\{\sum_{k+l=i+1,k,l\ge 1}a_ka_l+
\frac 1 {p^2+q^2+1}
\bigl[2\sum_{k+l=i,k,l\ge 0}a_k(pb_{l,x}+qc_{l,x})
\vspace{2mm}\\
\D  \qquad
+\sum_{k+l=i-1,k,l\ge 0}(b_{k,x}b_{l,x}+c_{k,x}c_{l,x})
\bigr]\bigr\},\ i\ge 1.
\ea  \]
Combined with \eqref{eq:RR3ofSZCE:ma-3rd-sh-so(3,R)},
a mathematical induction then shows that
the whole sequence of $\{a_i,b_i,c_i|\, i\ge 1\}$ is of differential function type.

Now, based on both the recursion relations in \eqref{eq:RR1_iofSZCE:ma-3rd-sh-so(3,R)} and \eqref{eq:RR2_iofSZCE:ma-3rd-sh-so(3,R)} and the structure of the spectral matrix $U$ in \eqref{eq:defofU:ma-3rd-sh-so(3,R)},
we introduce
\be V^{[m]}=
\lambda [(\lambda^m a)_+ U+(\lambda ^m b_x)_{+}e_2+(\lambda ^m c_x)_{+}e_3]
, \ m \ge 0,\ee
and see that the corresponding zero curvature equations
\begin{equation}
U_{t_m} -V^{[m]}_x +[U, V^{[m]}] = 0, \ m\ge 0,
\label{eq:ZCEforsh:ma-3rd-sh-so(3,R)}
\end{equation}
generate a hierarchy of soliton equations:
\begin{equation}
u_{t_m}= K_m=
\left[\ba {c}
b_{m,xx}
\vspace{2mm} \\
c_{m,xx}\ea
 \right]
 ,
\ m \ge 0,\label{eq:sh-so(3,R):ma-3rd-sh-so(3,R)}
\end{equation}
where are all local. In the next section, we are going to show that all those soliton equations are Liouville integrable.

\section{Bi-Hamiltonian structures}

\subsection{Hamiltonian structures}

To construct Hamiltonian structures, we apply the trace identity \eqref{eq:traceid:ma-3rd-sh-so(3,R)} (or more generally the variational identity
\eqref{eq:vid:ma-3rd-sh-so(3,R)}).
From the definition of $U$ and $W$ in \eqref{eq:defofU:ma-3rd-sh-so(3,R)} and \eqref{eq:W:ma-3rd-sh-so(3,R)},
it is direct to see that
\[ \frac {\part U}{\part \lambda} = \left[
\ba {ccc}
0&-q &-1\vspace{1mm}\\
q&0& -p \vspace{1mm}\\
1 &p & 0
 \ea \right],
  \
 \frac {\part  U}{\part p} = \left[
\ba {ccc}
0&0&0\vspace{1mm}\\
0&0& -\lambda  \vspace{1mm}\\
0&\lambda & 0
 \ea \right],\
 \frac {\part U}{\part q} = \left[
\ba {ccc}
0&-\lambda &0\vspace{1mm}\\
\lambda &0& 0 \vspace{1mm}\\
0&0& 0
 \ea \right],
  \]
and so,
we have
\be \ba {l}\textrm{tr}(W\frac {\part U}{\part \lambda} )= -2\lambda (p^2+q^2+1)a-2pb_x-2qc_x,\vspace{2mm}\\
\textrm{tr}(W\frac {\part U}{\part p} )=-2\lambda (\lambda pa+b_x)=2\lambda ^2 c,
\vspace{2mm}
\\
 \textrm{tr}(W\frac {\part U}{\part q} )=
 -2\lambda (\lambda qa+c_x)=
 -2\lambda ^2 b.
 \ea
 \nonumber
 \ee
 Now, in this case,
 the trace identity \eqref{eq:traceid:ma-3rd-sh-so(3,R)},
 i.e.,
 \[ \frac {\delta }{\delta u}
 \int \textrm{tr}(W\frac {\partial U}{\partial \lambda })\, dx=
 \lambda ^{-\gamma}\frac \partial {\partial \lambda}
 \lambda ^\gamma \textrm{tr}(W\frac {\partial U}{\partial u}),\
 u=\left [ \ba {c} p\vspace{2mm}\\  q\ea \right]
  \]
   presents
 \[
\frac{\delta}{\delta  u}
\int
[-\lambda (p^2+q^2+1)a-pb_x-qc_x]\, dx
=\lambda ^{-\gamma}\frac {\part }{\part \lambda }\lambda ^\gamma \left[ \ba {c}
\lambda ^2 c\vspace{2mm}\\
-\lambda ^2 b
\ea \right].
\]
Balancing coefficients of all powers of $\lambda $ in the equality tells
\be \frac{\delta}{\delta  u}
\int
[-(p^2+q^2+1)a_{0}]\, dx
=(\gamma +2)\left[ \ba {c}
c_0\vspace{2mm}\\
-b_0
\ea \right],
\label{eq:1stcomponentoftraceidentity:ma-3rd-sh-so(3,R)}\ee
and
\be
\frac{\delta}{\delta  u}
\int
[-(p^2+q^2+1)a_{m}-pb_{m-1,x}-qc_{m-1,x}]\, dx
=(\gamma - m+2)\left[ \ba {c}
c_m\vspace{2mm}\\
-b_m
\ea \right], \ m\ge 1.
\label{eq:mthcomponentoftraceidentity:ma-3rd-sh-so(3,R)}
\ee
Checking a particular case in \eqref{eq:1stcomponentoftraceidentity:ma-3rd-sh-so(3,R)} yields $\gamma=-1$, and thus we obtain
\be \frac{\delta {\cal H}_{m}}{\delta  u}
=\left[ \ba {c}
c_{m}\vspace{2mm}\\
-b_{m}
\ea \right], \ m\ge 0,
 \ee
where \be
{\cal H}_{0}=\int (- \sqrt{p^2+q^2+1}\,) \, dx,\
{\cal H}_{1}=\int \frac {qp_x-pq_x}{\sqrt{p^2+q^2+1}\,(\sqrt{p^2+q^2+1}+1)} \, dx, \label{eq:1H_m:ma-3rd-sh-so(3,R)} \ee
and
\be
{\cal H}_{m+1}=\int
\frac {(p^2+q^2+1)a_{m+1}+pb_{m,x}+qc_{m,x}}{m}\, dx ,\ m\ge 1.\label{eq:2H_m:ma-3rd-sh-so(3,R)}
\ee
Here ${\cal H}_1$ was directly computed, since when $m=1$, the coefficient on the right hand side of \eqref{eq:mthcomponentoftraceidentity:ma-3rd-sh-so(3,R)} is zero.
It then follows that the soliton hierarchy \eqref{eq:sh-so(3,R):ma-3rd-sh-so(3,R)} has the first Hamiltonian structures:
 \begin{equation}
  u_{t_m}=K_m= \left[\ba {c}
b_{m,xx}
\vspace{2mm} \\
c_{m,xx}\ea
 \right]
 =J\left[\ba {c}
c_{m}
\vspace{2mm} \\
-b_{m}\ea
 \right]
 =  J\frac {\delta {{\cal  H}}_m} {\delta  u},\ m\ge 0,
\end{equation}
where the Hamiltonian operator is defined by
\be J=\left[ \begin{matrix}
0  & -\partial ^2 \vspace{2mm}
\\ \partial ^2 & 0
\end{matrix}\right],\label{eq:J:ma-3rd-sh-so(3,R)} \ee
and the Hamiltonian functionals, by \eqref{eq:1H_m:ma-3rd-sh-so(3,R)} and \eqref{eq:2H_m:ma-3rd-sh-so(3,R)}.

The obtained functionals $\{{\cal H}_m\}_0^\infty$ generate
an infinite sequence of conservation laws, not being of differential polynomial type, for each member in the counterpart hierarchy \eqref{eq:sh-so(3,R):ma-3rd-sh-so(3,R)}.
 We point out that conservation laws of differential polynomial type
  can be computed
  systematically through B\"acklund transformations (see, e.g., \cite{Satsuma-PTH1974,WadatiSK-PTP1975}),
   from a Riccati equation
  generated from the underlying spectral problems (see, e.g., \cite{CasatiDO-JGP2008,AlbertyKS-PD1982})
  or by using computer algebra systems (see, e.g., \cite{HeremanAEHH-book2009}).

\subsection{Bi-Hamiltonian structures}

It is now a direct
but lengthy computation
to show by computer algebra systems that
 $ J$ defined by \eqref{eq:J:ma-3rd-sh-so(3,R)} and
  \begin{equation}
  M=J\Psi=\Psi^\dagger J=
    \left[ \begin{matrix} \part ^3 - \part ^2 \tilde q \part ^{-1}\tilde q \part ^2 & \part ^2 \tilde q \part ^{-1}\tilde p \part ^2 \vspace{2mm}
\\ \part ^2  \tilde p \part ^{-1} \tilde q\part ^2 & \part ^3 -\part ^2 \tilde p \part ^{-1} \tilde p \part ^2
\end{matrix}\right]
  ,\label{eq:defofM:ma-3rd-sh-so(3,R)}
\end{equation} constitute a Hamiltonian pair (see \cite{Magri-JMP1978,FuchssteinerF-PD19812} for examples),
where $\Psi$ is defined as in \eqref{eq:RR1_iofSZCE:ma-3rd-sh-so(3,R)} and $\Psi^\dagger$ denotes the conjugate operator of $\Psi$.
Consequently,
  any linear combination $ N$ of $ J$ and $ M$ satisfies
  \be
  \int  K ^T  N'( u) [ N S] T  \, dx +\textrm{cycle}( K, S, T) =0
  \ee
  for all vector fields $ K$, $ S$ and $ T$.
This implies that
the operator $ \Phi=\Psi^\dagger$
is hereditary (see \cite{Fuchssteiner-NA1979} for definition), i.e.,
it satisfies
 \be  \Phi'( u)[ \Phi  K] S -  \Phi  \Phi'( u) [ K] S =   \Phi'( u)[ \Phi  S] K
 - \Phi \Phi'( u) [ S] K \label{eq:hereditaryproperty:ma-3rd-sh-so(3,R)}
  \ee
 for all vector fields $ K$ and $ S$.
The condition \eqref{eq:hereditaryproperty:ma-3rd-sh-so(3,R)} for the hereditariness is equivalent to
\be L_{\Phi K}\Phi =\Phi L_K\Phi  \ee
where $K$ is an arbitrary vector field. The Lie derivative $L_K\Phi$ here is
defined by
\[ (L_K\Phi )S= \Phi [K,S]-[K,\Phi S],\]
with $[\cdot,\cdot]$ being the Lie bracket of vector fields:
\[ [K,S]=K'(u)[S]-S'(u)[K],\]
where $K'$ and $S'$ denotes their Gateaux derivatives.

Note that an autonomous operator $\Phi=\Phi(u,u_x,\cdots) $
is a recursion operator of a given evolution equation $u_t=K=K(u)$ if and only if $\Phi$ needs to satisfy
\be L_{K}\Phi=0. \ee
It is easy to see that  the operator $\Phi=\Psi^\dagger$ satisfies \be
L_{K_0}\Phi=0,\ \ \textrm{where}\ \ K_0=\left[\ba {c} \Bigl(\frac q{\sqrt{p^2+q^2+1}}\Bigr)_{xx} \vspace{2mm}\\ -\Bigl(\frac p{\sqrt{p^2+q^2+1}}\Bigr)_{xx} \ea \right],\ \ee and thus
\[ L_{K_m}\Phi = L_{\Phi K_{m-1}}\Phi  =\Phi L_{K_{m-1}}\Phi=\cdots=\Phi^m L_{K_0}\Phi =0,\ m\ge 1,\]
where the $K_m$'s are defined by \eqref{eq:sh-so(3,R):ma-3rd-sh-so(3,R)}.
This implies that the operator $\Phi=\Psi^\dagger$ is a common hereditary recursion operator for the counterpart soliton hierarchy \eqref{eq:sh-so(3,R):ma-3rd-sh-so(3,R)}. We point out that there are also a few direct symbolic algorithms
for computing recursion operators of nonlinear partial differential equations
by
computer algebra systems (see, e.g., \cite{BaldwinH-IJCM2010}).

It now follows that all members, except the first one, in the counterpart soliton hierarchy \eqref{eq:sh-so(3,R):ma-3rd-sh-so(3,R)}
are bi-Hamiltonian:
\be
u_{t_m}=K_m=J\frac {\delta {\cal H}_{m}}{\delta u}=M\frac {\delta {\cal H}_{m-1}}{\delta u},\ m\ge 1, \ee
where $J,M$ and ${\cal H}_m$'s are defined by
\eqref{eq:J:ma-3rd-sh-so(3,R)},
\eqref{eq:defofM:ma-3rd-sh-so(3,R)}, \eqref{eq:1H_m:ma-3rd-sh-so(3,R)} and
\eqref{eq:2H_m:ma-3rd-sh-so(3,R)}.
 Therefore, the counterpart hierarchy \eqref{eq:sh-so(3,R):ma-3rd-sh-so(3,R)} is Liouville integrable, i.e., it possesses infinitely many commuting symmetries and conservation laws.
  Particularly,
we have the Abelian symmetry algebra:
\be [ K_k, K_l]= K_k'( u)[ K_l]-
 K_l'( u)[ K_k]
=0,\ k,l\ge 0, \ee
and the two Abelian algebras of conserved functionals:
\be \{ {\cal H}_k, {\cal H}_l\}_{ J}=
\int \bigl(\frac {\delta { {\cal H}}_k}{\delta  u}\bigr)^T J \frac {\delta { {\cal H}}_l}{\delta u}\,dx
=0,\ k,l\ge 0,\ee
and
\be \{ {\cal H}_k, {\cal H}_l\}_{ M}=
\int \bigl(\frac {\delta { {\cal H}}_k}{\delta  u}\bigr)^T M \frac {\delta { {\cal H}}_l}{\delta u}\,dx
=0,\ k,l\ge 0.\ee

The first nonlinear bi-Hamiltonian integrable system in the counterpart soliton
hierarchy \eqref{eq:sh-so(3,R):ma-3rd-sh-so(3,R)} is as follows:
\be
u_{t_1}=\left[\ba {c} p\vspace{2mm} \\
q \ea \right]_{t_1}=K_1=
-\left[\ba {c} \Bigl( \frac {p_x}{(p^2+q^2+1)^{\frac 32 }}\Bigr)_{xx} \vspace{2mm} \\
\Bigl( \frac {q_x}{(p^2+q^2+1)^{\frac 32 }}\Bigr)_{xx} \ea \right] =J
\frac {\delta {\cal H}_{1}}{\delta u}=M\frac {\delta {\cal H}_{0}}{\delta u}
.
\ee
This is a different system from the WKI system of nonlinear soliton equations presented in \cite{WadatiKI-JPSJ1979}.

\section{Concluding remarks}

Based on the real matrix loop algebra $\widetilde {\textrm{so}}(3,\mathbb{R})$,
we
formulated a spectral problem by the same linear combination of basis vectors as the WKI one, and
introduced a counterpart of the WKI soliton hierarchy by the zero curvature formulation, whose soliton equations are  of differential function type but not
 of differential polynomial type. By the trace identity,
the resulting counterpart soliton hierarchy has been shown to be bi-Hamiltonian and so Liouville integrable.

The real Lie algebra
of the special orthogonal group,
$\textrm{so}(3,\mathbb{R})$, is not isomorphic to the real Lie algebra $\textrm{sl}(2,\mathbb{R})$ over $\mathbb{R}$, and
thus the newly presented soliton hierarchy \eqref{eq:sh-so(3,R):ma-3rd-sh-so(3,R)} and the WKI soliton hierarchy \cite{WadatiKI-JPSJ1979}
are not gauge equivalent over $\mathbb{R}$.
The main difference between the WKI soliton hierarchy and the counterpart soliton hierarchy is that
the second Hamiltonian operators are different, which are
\[
M=\left[ \begin{matrix}
\frac 14 \part ^2 \bar p\part ^{-1}\bar p\part ^2 & \frac 12 \part ^3 -\frac 14 \part ^2 \bar p\part ^{-1}\bar q\part ^2 \vspace{2mm}
\\ \frac 12 \part ^3  -\frac 14 \part ^2 \bar q\part ^{-1}\bar  p\part ^2 & \frac 14 \part ^2 \bar q\part ^{-1} \bar q \part ^2
\end{matrix}\right]
,\]
and
\[ M=\left[ \begin{matrix}
\part ^3 - \part ^2 \tilde q \part ^{-1}\tilde q \part ^2 & \part ^2 \tilde q \part ^{-1}\tilde p \part ^2 \vspace{2mm}
\\ \part ^2  \tilde p \part ^{-1} \tilde q\part ^2 & \part ^3 -\part ^2 \tilde p \part ^{-1} \tilde p \part ^2
\end{matrix}\right],\]
where
\[ \bar p=\frac p {\sqrt{pq+1}},\ \bar q=\frac q {\sqrt{pq+1}},\]
and
\[\tilde p =\frac p {\sqrt{p^2+q^2+1}},\ \tilde q =\frac q {\sqrt{p^2+q^2+1}}.\]
They constitute two Hamiltonian pairs
with
the first Hamiltonian operators
\[J=\left[ \begin{matrix}
0  & \part ^2 \vspace{2mm}
\\ -\part ^2 & 0
\end{matrix}\right],\
J=\left[ \begin{matrix}
0 &  -\partial ^2\vspace{2mm}
\\ \partial ^2 & 0
\end{matrix}\right],
\]
and
 generate two different hereditary recursion operators:
 \[
\Phi=
\left[
\ba {cc}   \frac 12 \part -\frac 14 \part ^2 \frac 1 {\bar p} \part ^{-1} \frac 1 {\bar q} &
-\frac 14 \part ^2 \frac 1 {\bar p} \part ^{-1}\frac 1{\bar p}
\vspace{2mm}\\
 \frac 14 \part ^2 \frac 1 {\bar q} \part ^{-1} \frac 1 {\bar q} &
-\frac 12 \part  +\frac 14  \part ^2 \frac 1 {\bar q} \part ^{-1} \frac 1 {\bar p}
\ea \right]
,
 \]
 and
\[\Phi=
\left[
\ba {cc}  -\part ^2 \frac 1 {\tilde q} \part ^{-1}\frac 1 {\tilde p} & \part
-\part ^2 \frac 1 {\tilde q} \part ^{-1}\frac 1 {\tilde q}
\vspace{2mm}\\
-\part  +\part ^2 \frac 1 {\tilde p} \part ^{-1} \frac 1 {\tilde p} &
  \part ^2 \frac 1 {\tilde p} \part ^{-1} \frac 1 {\tilde q}
\ea \right],\]
respectively.

We remark that for a bi-Hamiltonian soliton hierarchy, one can make a kind of nonholonomic constraint
\[ M\left[\ba {c} f\vspace{2mm}\\ g \ea \right]
=K_0,\]
by using the first vector field $K_0$ and the second Hamiltonian operator $M$.
    Starting with such functions $f$ and $g$, normally being nonlocal, and applying the first Hamiltonian operator $J$, one can introduce
a so-called negative system of soliton equations
\[ u_{t_{-1}}=J \left[\ba {c} f\vspace{2mm}\\ g \ea \right],
\]
and step by step, the whole negative hierarchy, which still has zero curvature representations similar to the ones for a given soliton hierarchy.
However, in our case associated with so(3,$\mathbb{R}$),
the nonholonomic constraint itself
 defines an integro-differential system for $f$ and $g$, which goes beyond our focused scope.

We also point out that there has recently been a growing interest in soliton hierarchies generated from spectral problems associated with non-semisimple Lie algebras.
Various examples of bi-integrable couplings and tri-integrable couplings offer inspiring insights into the role they plays in classifying
  multi-component integrable systems \cite{MaMZ-GJMS2012}.
 Multi-integrable couplings do bring diverse structures on recursion operators in block matrix form \cite{Ma-book2010,MaMZ-GJMS2012}.
 It is significantly important in helping to understand
 essential properties of integrable systems
 to explore more mathematical structures behind integrable couplings.

It is known that there exist Hamiltonian structures for the perturbation equations \cite{MaF-CSF1996,Sakovich-JNMP1998}, but
it is not clear
how one can generate Hamiltonian structures for general integrable couplings
\cite{MaG-MPLB2009,Ma-PLA2003}.
There is no guarantee that there will exist non-degenerate bilinear forms required in the variational identity
on the underlying non-semisimple matrix Lie algebras.
  It is particularly interesting to see
  when Hamiltonian structures can exist for bi- or tri-integrable couplings \cite{Ma-CAMB2012,MaZM-EAJAM2013,MaMZ-IJNSNS2013},
  based on algebraic structures
  of non-semisimple matrix loop algebras.
  A basic question in the Hamiltonian theory of integrable couplings is
whether there is any Hamiltonian structure
for
the bi-integrable coupling
\[u_t=K(u),\ v=K'(u)[v],\ w_t=K'(u)[w],\]
  where $K'$ is the Gateaux derivative.

  \vskip 2mm

\noindent{\bf Acknowledgements:}
The work was supported
 in
part by NSF under the grant DMS-1301675,
 NNSFC under the grants
  11371326, 11271008, 61070233, 10831003 and 61072147,
 Chunhui Plan of the Ministry of Education of China,
 Zhejiang Innovation Project of China (Grant No. T200905),
 and
 the First-class Discipline of Universities in Shanghai and the Shanghai Univ. Leading Academic Discipline Project (No. A.13-0101-12-004).
  The authors are also grateful to E. A. Appiah, X. Gu,  C. X. Li, M. McAnally, S. M. Yu and W. Y. Zhang
  for their stimulating discussions.


\end{document}